# *Polymer fiber mats provide white and full-color tunable lasing*

Alina Szukalska*[1], Kinga Halicka-Stępień[1], Jastin Popławski[1], Katarzyna Kołodzińska[1], Michał Majkowski[2], Joanna Cabaj[1], Lech Sznitko*[1]

[1] *Wroclaw University of Science and Technology, Soft Matter Optics Group, Department of Chemistry, Wybrzeże Wyspiańskiego 27, 50-370 Wroclaw, Poland.*

[2] *University of Wroclaw, Faculty of Biotechnology, Plac Uniwersytecki 1, 50-137 Wroclaw, Poland*

Corresponding authors: lech.sznitko@pwr.edu.pl, alina.szukalska@pwr.edu.pl

**Abstract:**

Random Lasing has become highly advantageous for achieving white laser emission. Utilizing multiple light scattering in disordered media allows for cost-effective and flexible designs, as well as the seamless integration of multiple multicolor gain materials. However, generating light that closely meets the D65 illumination standard remains challenging due to energy transfer between chromophores. To address this issue, effective spatial separation of gain media is required. Here, we present a solution to the problem using hierarchically designed polymer fiber mats with spatially distributed red, green, and blue (RGB) laser dyes to achieve controlled and tunable multicolor and white lasing. While electrospun fibers are well-studied, their use for multicolor random lasing is demonstrated here for the first time. Our results demonstrate outstanding spectral tunability, aligning well with industrial color standards such as sRGB, Adobe RGB, and DCI-P3. The white lasing closely matches D-series white illuminants, including D65, D55, and D75 standards, with coordinate deviations as low as approximately 0.2 to 3%, surpassing previously reported values for organic, hybrid, and inorganic systems. By limiting energy transfer, our approach enables precise lasing control, demonstrating that fiber mats are versatile for next-generation random lasing applications. These findings advance technologies in Li-Fi, laser displays, optical sensors, switches, and modulators, bringing tunable lasers closer to commercialization.

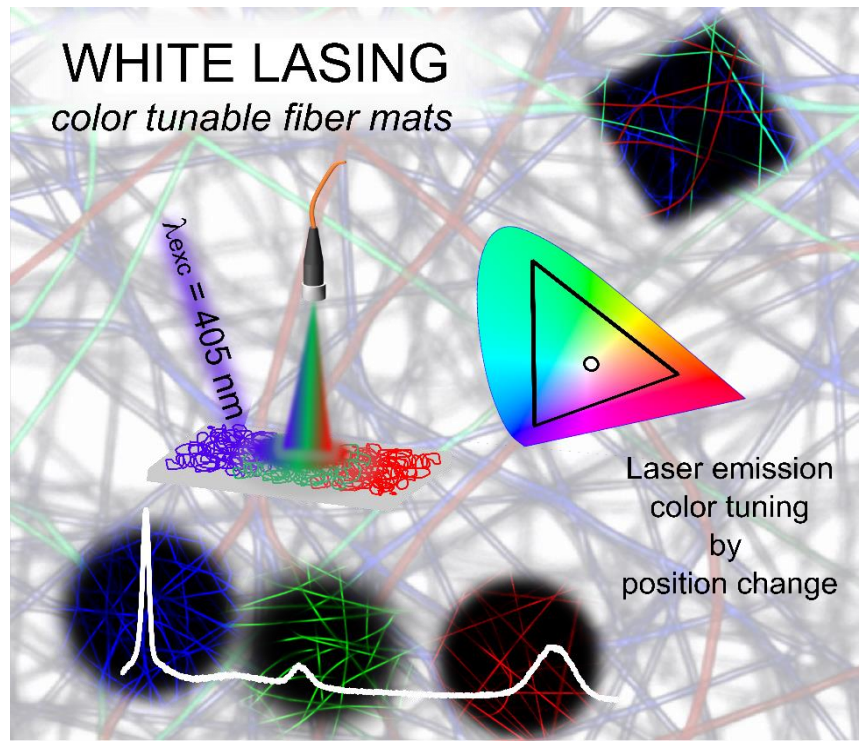

**Introduction:**

Random lasers (RLs) have gained attention for their unique properties, offering new opportunities in displays, lighting, and imaging systems [1–3]. Unlike conventional lasers using Fabry-Perot mirrors, RLs rely on multiple light scattering within a disordered medium for photon flux amplification. This approach offers flexibility, allowing various materials, such as quantum dots, liquid crystals, or saturated solutions, to act as scatterers [4–8]. Therefore, RLs are considered versatile and cost-effective systems. Additionally, RLs can use multiple gain-providing compounds, like laser dyes emitting at different wavelengths, to produce multicolor, tunable emissions from a single device. This capability drives innovations in technologies such as Red + Green + Blue (RGB) RLs for achieving white light, which is crucial for advanced displays and lighting systems [9,10]. One notable development is Light Fidelity (Li-Fi) [11–13], which combines white emission with data transmission, offering an increasing data transmission speed of up to two orders of magnitude compared to Wi-Fi. Furthermore, laser displays provide a 70% wider color range than other systems. While the first commercial laser projectors are already available, ongoing research aims to scale down the technology and reduce costs by identifying optimal materials [14–16].

Therefore, white lasing currently is a highly desirable phenomenon; however, reaching a pure white color that meets the D65 white illumination standard remains challenging [17]. It is due to difficulties such as achieving population inversion across multiple dyes, balancing RGB lasing bands, ensuring compatible absorption ranges to use a single excitation source, and addressing energy transfer processes. Avoiding particularly troublesome Förster Resonance Energy Transfer (FRET) [18,19] can be obtained by controlling the material composition and spatially separating the gain media. A simple and cost-effective method, known in the literature as electrospinning, appears as an excellent approach for addressing energy transfer issues and achieving white lasing [20,21]. Electrospinning involves applying a high-voltage electric field to a polymer solution or melt, forming fine fibers with diameters in the nanometer/micrometer range [22–25]. The technique is very straightforward, cost-effective, and scalable [26,27] It can be applied to various materials, such as polymers, biopolymers, and organic-inorganic hybrid systems [25]. The versatility of this process allows the development of systems with tailored optical properties [21,28,29], including RL. For example, in [30], nanofibers doped with the gain molecule (Rhodamine 6G) are demonstrated to provide both Amplified Spontaneous Emission and lasing. This makes electrospun nanofibers promising for applications requiring both coherent and incoherent light sources, with the ability to tune the emission mechanism. This extends to Random Lasing, where their disordered, porous structure enhances light scattering. Moreover, an important advantage of electrospun RLs is the ability to control lasing modes by optimizing non-uniform pump patterns, enabling selective amplification or suppression of specific modes [31]. This restraint is crucial for advanced optical devices used in photonic circuits [32] or for sensing [33]. The precise manipulation of lasing modes can also be realized by controlling the fiber geometry [30]. Furthermore, strategies for tuning emission include manipulating fiber arrays, diameters, and refractive index changes, offering fine adjustment over lasing performance [34–36].

Building on the versatility of electrospinning, Zhou et al. [37] demonstrated an interesting gradient fiber mat (FM) idea by using a cone-shaped rotating collector. The structural variation is achieved by controlling the polymer jet's travel distance and the collector's surface velocity, affecting fiber morphology such as diameter, porosity, and alignment. This technique aims to replicate natural tissue structures, enhancing scaffolds' mechanical properties and biological functionality for biomedical applications.

Inspired by Zhou et al.'s paper, we designed multicolor FMs with one, two, or three dyes arranged in a gradient. This allows for control of the lasing, transitioning smoothly from one dye's emission to another while also blending the colors where they overlap. This precise tuning offers a fresh concept to adjust optical responses, making produced devices attractive for RLs and other photonic devices, like optical sensors, switches, or modulators. We systematically explore single-, through dual, to triple-color FM devices, optimized for efficient, controlled polychromatic lasing, leading to highly desirable white lasing. Compared to standard RGB ranges, our results show that our devices offer much wider tuning potential than required for industrial applications such as sRGB, AdobeRGB, and DCI-P3, and are competitive with the latest standard Rec 2020. By comparing achieved results with white standard illuminants from the D series, we highlight the technological potential of produced mats, which outperform previously reported inorganic, hybrid, and organic materials [38–41]. Namely, our results align with D50, D55, and D65 standards by 0.2 - ~3%, whereas those reported in the literature often deviate from the benchmarks by around 10 to 20%, considering the x and/or y coordinates [39,40]. Finally, the limitation of FRET is documented through energy transfer studies and fluorescence decay time measurements in the produced devices, showing the controllable design of RGB, light-emitting systems.

## Results and discussion

### *Electrospinning process and SLFMs*

Investigated FMs were fabricated as described in the Methods section. Typically, spinning time is performed over 10 - 60 minutes [42,43] to ensure uniform layer formation. However, to achieve laser emission from multiple-layer FMs with a hierarchical design, the spinning time was reduced (to 0.5 - 3 min) compared to previous studies. This leads to less dense fiber packing and allows FMs to maintain optical transparency. As a result, the light can pass through multiple layers, enabling each to contribute to the collective laser emission. Moreover, it facilitates gradient design for multicolor laser emission and, ultimately, white lasing. The process of electrospinning is schematically shown in Fig. 1 (a).

***Fig. 1: Electrospinning setup, gradient morphologies, and characterization of SLFMs.***

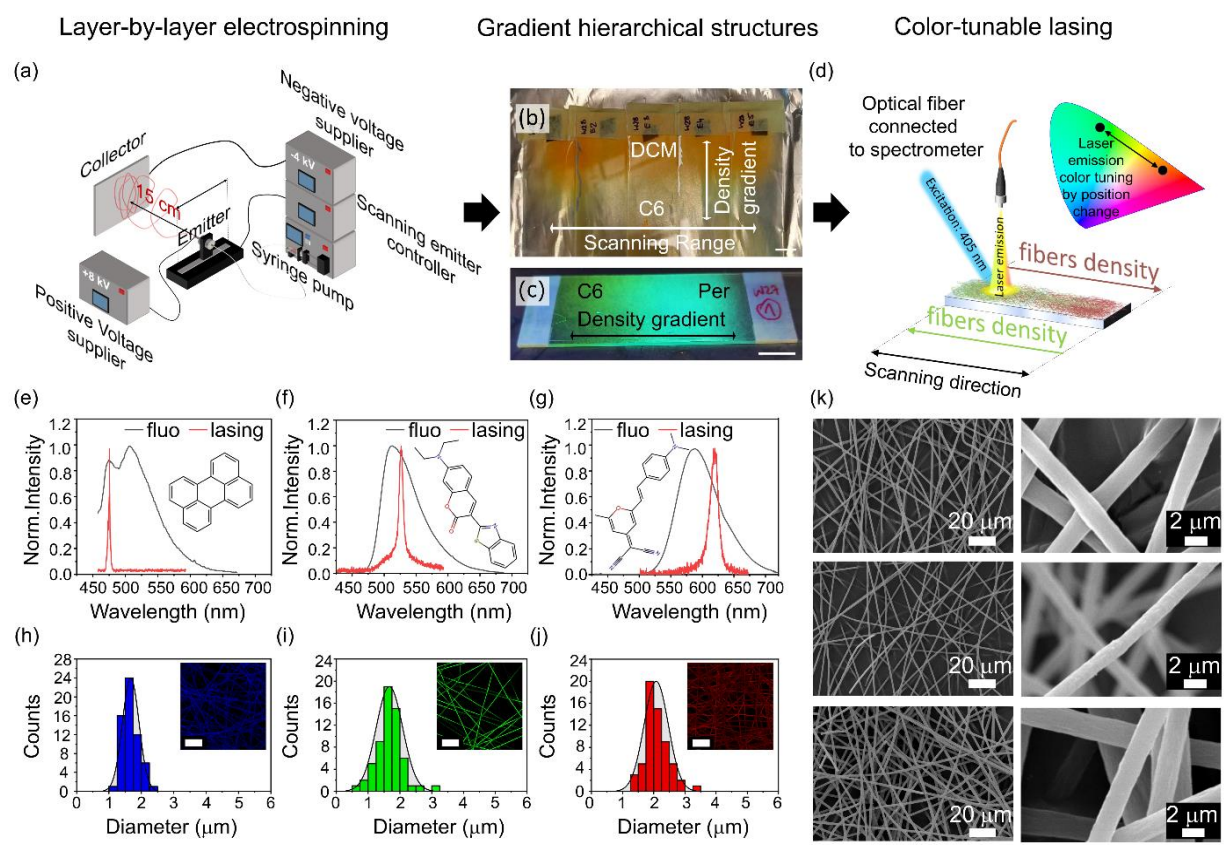

***Fig. 1*** *Schematic representation of electrospinning setup (a). Photographic demonstration of the gradient distributions in DCM-C6 (b) and Per-C6 (c) samples under UV illumination (scale bar: 1cm). The idea of color tunability using a gradient of fibers' density (d). PL and RL spectra for Per (e), C6 (f), and DCM (g) embedded in SLFMs; insets show the dyes' chemical structures. Histograms of fiber diameters measured from confocal images obtained for Per (h), C6 (i), and DCM (j) loaded SFLMs; insets show confocal images. SEM images for Per, C6, and DCM containing SLFMs (k), left column – smaller magnification, scale bar 20μm, right column – larger magnification, scale bar 2μm.*

Typically, fibers in the electrospinning process are spun over the area with the most suitable electric potential. For samples already containing a fiber layer, subsequent layers are deposited in regions where the previous layer does not screen the potential. More details are available in the Electronic Supplementary Information (ESI), section S1 "Fiber's density studies", Figure S1. In our experiments, the area covered by fibers typically had an elliptic shape (Fig.1 (b)) of ~12cm on the long axis and ~4cm on the short axis. The most significant change in fiber density is observed at the edge of the coated area, typically ~1 cm wide. By properly placing a substrate on the collector, it is possible to harvest the non-woven with a controlled fiber density gradient on the resulting sample.

For double-layer FMs (DLFM), we consistently observed a shift of approximately ∼ *1.0 – 1.5cm* between the two layers. Thus, by properly fixing the glass substrate on the collector, it is possible to control the morphology and fiber density relative to the sample position (c.f. Fig. 1 (b) and (c)). More details about the composition and fabrication parameters of the obtained FMs are gathered in section S2 in the Table. S1. Such a sample design offers emission color tunability controlled by the change in the position of optical excitation, as schematically shown in Fig. 1 (d). To allow control over a wide range of colors, three RGB dyes previously examined by our group [44,45] were selected: DCM for red (R), C6 for green (G), and Per for blue (B) emission (full names in the Methods section). All the mentioned compounds show efficient fluorescence and the ability to achieve population inversion.

We began our examination with the simplest possible system, focusing on Single-Layer Fiber Mats (SLFMs), which we used as the starting point for more complex configurations presented further. Fig. 1 (e)-(g) shows the photoluminescence (PL) and RLng spectra collected from three separate SLFMs with *Per*, *C6*, and *DCM* dyes, respectively. Typically, RL emission is much narrower than PL and is characterized by energy threshold conditions. These can be estimated using light-in-light-out (Li-Lo) plots and by monitoring the full width at half maximum (FWHM) of the collected emission, as shown in section S3, Fig. S2 (a)-(c). The threshold energy values are the lowest for *C6* dye-doped FM, measuring approximately 140 µJ/pulse (22 kW). The *DCM* is exhibiting a moderate threshold level, ~500 µJ (78 kW), and the *Per*-doped SLFM is the hardest to pump, with a threshold level of around 1700 µJ (266 kW). For more details, see the Tab. S2 in ESI. Closer inspection of the FM's morphologies using confocal (insets in Fig. 1. h-j) and scanning electron microscope (Fig. 1k) (SEM) (see Methods section) revealed their sizes and shapes. In all cases, the fibers have round cross-sections with diameters ranging from approximately 1.0 to 2.5 µm. SEM images show smooth structures with minimal surface roughness. The diameters are sufficiently large to support a multimodal waveguiding effect, consistent with the findings of Camposeo et al. [35].

The morphologies of the obtained hierarchical fiber mats are rather typical for the electrospinning process of PMMA polymer with smooth surfaces and round cross-sections. The size of all investigated fibers is somewhat similar, showing a minimal impact of the used dye dopant (See Fig. 1 (h-j)), spinning time, and order in deposition hierarchy.

As previously described [44] such a hierarchical design can strongly reduce the (FRET) between different dyes due to their physical separation, allowing easier control over sample composition to reach the desired color of laser emission. Fluorescence lifetime imaging microscopy (FLIM) evidenced this feature. PL intensity decay fitting parameters are shown in ESI, section S4 in Fig. S3-S4, while the estimations of PL lifetimes and FRET are shown in Fig. S5. All details are described in ESI in the FLIM and FRET estimation section. It can be seen that FRET plays a marginal role when amplitude-weighted average lifetimes are considered. But if we include errors in the PL lifetime estimation, there is a non-zero probability that this effect is entirely suppressed.

### *DLFMs – comparable PL and RL performance and controlled color tunability*

We demonstrated that all fabricated and optimized SLFM structures enable efficient RL performance. Building on these results, we next explore DLFWMs. We present a controlled and straightforward approach to achieve tunable, multicolor lasing by introducing dual-dye systems. DLFMs were characterized according to PL and RL color tunability, as described in the Methods section. The RL measurements were performed with an excitation beam characterized by energy equal to 4 mJ to ensure that all dyes would lase. At first glance, it is evident that in the *Per-C6* DLFWM, the PL and RL spectra (Fig. 2(a), (b)) evolve as the excitation position moves along the x-direction (approximately *4 cm* from one side to the other), with a gradual shift in the peak dominance corresponding to the transition from *C6*-dominated regions to those where *Per* emission becomes more pronounced.

**Figure 2: Multicolor PL and RL performance of position-tunable DLFMs**

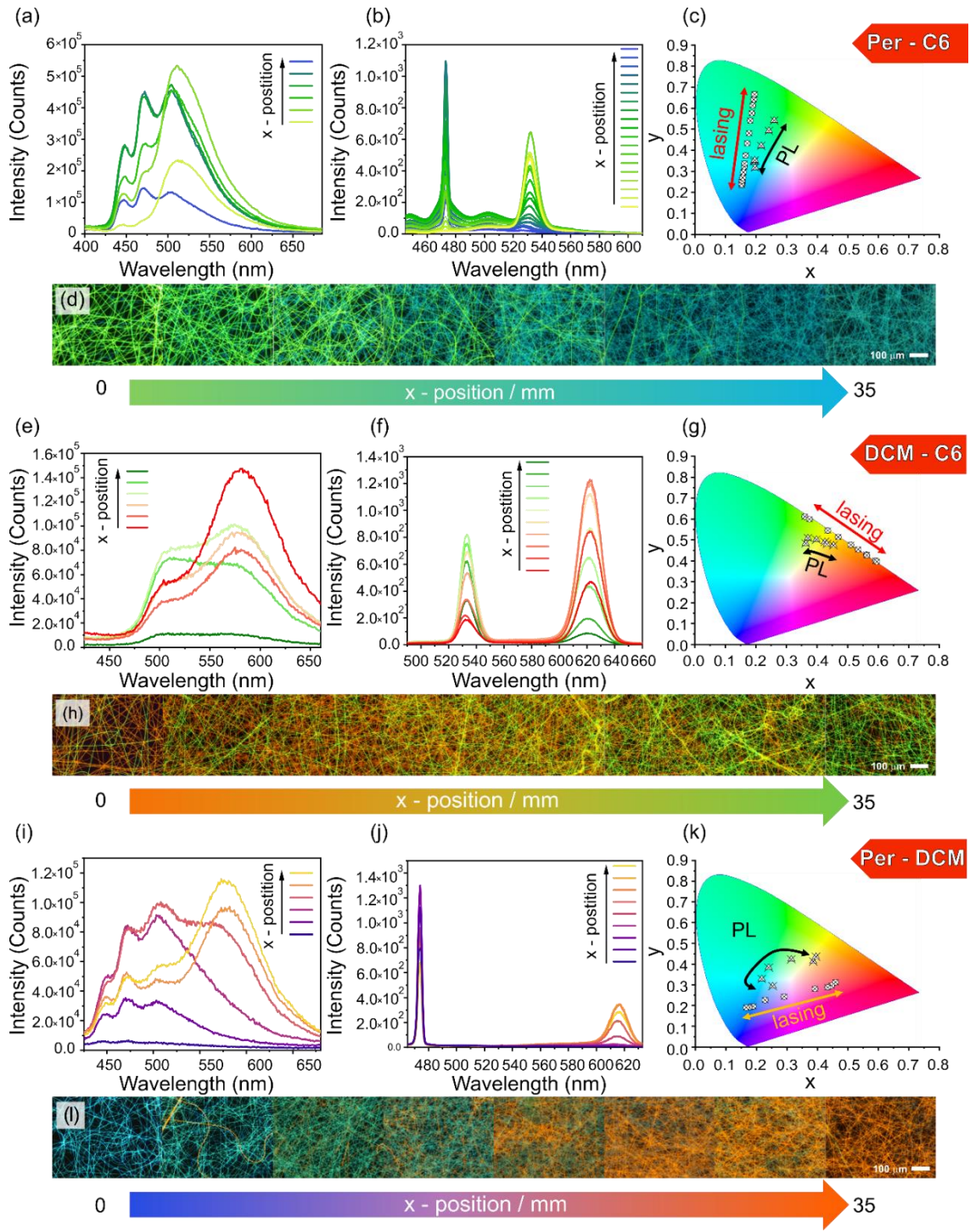

***Fig. 2*** *Per-C6 DLFM PL (a), RL (b) spectra change concerning excitation position and their coordinates on CIE1931 chromatic diagram (c), a set of 8 microscopic images showing the morphological changes for different x-positions ranging from 0 to 35 mm with 5 mm step (d). DCM-C6 DLFM PL (e), RL (f) spectra change concerning excitation position and their coordinates on CIE1931 chromatic diagram (g), a set of 8 microscopic images showing the morphological changes for different x-positions ranging from 0 to 35 mm with a 5 mm step (h). Per-DCM DLFM PL (i), RL (j) spectra change concerning excitation position and their*

*coordinates on CIE1931 chromatic diagram (k), a set of 8 microscopic images showing the morphological changes for different x-positions ranging from 0 to 35 mm with a 5 mm step (l).*

In both cases, at the initial x-position (See Fig. 1c), the emission is dominated by the *C6* dye. As the sample is scanned along the x-direction, the peak contribution of *Per* rises in both collective PL and RL spectra. Simultaneously, the *C6* emission decreases, resulting in a change in the observed color. Fig. 2 (c) compares the color tunability of PL and RL on the CIE1931 diagram for the *Per-C6* sample. RL arises from stimulated emission, which selectively amplifies specific wavelengths based on the gain and feedback conditions, resulting in sharp, well-defined peaks in the emission spectrum. It corresponds to highly saturated colors, which place the emission points closer to the edges of the CIE diagram. In contrast, PL originates from spontaneous emission, where photons are emitted across a broader range of wavelengths, producing wider spectra and colors positioned closer to the central part of the chromaticity triangle. This distinction is evident in our case, where the lasing process achieves a broader range of accessible colors compared to PL. As a result, the RL emission can be tuned from vivid sky blue, through turquoise to mint green. In contrast, the PL emission allows for colors ranging from light blue to light green, over a much narrower range. Fig. 2 (d) shows microscopic images of the mentioned sample, taken every *5 mm* along the x-position using a fluorescence microscope. We estimated the densities of fibers using *Fiji* software from the presented set of images. Results are demonstrated in section S5 Fig. S6 in Electronic Supporting Information (ESI) showing the total fiber area (TFA) in the observed region. The area occupied by Per fibers is increasing from approximately $0.13\times10^5$ to $2.22\times10^5 \mu m^2$ while the area occupied by C6 fibers is decreasing from $1.70\times10^5 \mu m^2$ to 0 (with the total observed area under the microscope being $2.57\times10^5 \mu m^2$).

A similar behavior is observed for the *DCM-C6* sample. At the initial position, the *DCM* contribution dominated both the PL and RL spectra. However, as the sample is moved along the x-axis, the intensity of the *C6* band gradually increases, eventually becoming the dominant one in both spectra. (c.f., Fig. 2 (e) and (f)). Similarly, the range of available colors was much broader for RL, ranging from red, through orange and yellow, up to apple green (see Fig. 2. (g)). Since the first layer (C6) was spun for only 1 minute, it did not sufficiently shield the electric potential for the subsequent layer. As a result, the second layer was deposited with much smaller density differences compared to the previous case. As shown in Fig. S6, the visible area covered by *C6* fibers increases from $0.20 \times 10^5$ to $1.46 \times 10^5 \mu m^2$ (for x = *30mm*) and then rapidly decreases. For *DCM* fibers, the highest TFA was reported for x = 5*mm* ($1.29 \times 10^5 \mu m^2$) and decreased to $0.20 \times 10^5 \mu m^2$ at x = 40*mm*. The TFA for both types of fibers remains nearly constant between x-positions of 5 mm and 35 mm, as shown in section S5, Fig. S6.

Finally, similar measurements were conducted on the final DLFM sample containing *Per* and *DCM*. As shown in Fig. 2 (i), at the initial position, the PL spectrum is dominated by weak *Per* emission due to the small number of fibers. As the sample is shifted, the intensity of the *Per* emission increases and the *DCM* band gradually appears in the PL spectrum. Finally, the *DCM* band dominates the emission while the *Per* emission vanishes. The *x* and *y* coordinates on the CIE1931 diagram depend on the intensity (not only spectrum shape); therefore, the color change is not straightforward, as in previous cases. The PL color changes from pastel blue, aquamarine, green light, and green to light yellow, bypassing the region of white. On the other hand, for RL, the color palette is much broader, spanning a straight line on the diagram from saturated blue to rose. The *Per* fibers are initially deposited for 3 minutes, resulting in almost uniform coverage across x positions ranging from *10 mm* to *35 mm*, with TFA values between 1.5 and $1.7 \times 10^5 \mu m^2$. At lower x positions, such as *5 mm* ($1.32 \times 10^5 \mu m^2$) and *0 mm* ($0.65 \times 10^5 \mu m^2$), the covered areas are smaller, but *Per* fibers are still present. In contrast, *DCM* fibers occupy positions where the amount of *Per* fibers is lower. For x-positions between 0 and 10 mm, the TFA ranges from $1.86 \times 10^5$ to $2.03 \times 10^5 \mu m^2$. At x = *15 mm*, a sudden drop in the covered area is observed, and by x = *35 mm*, no *Per* fibers are visible (c.f., section S5. Fig. S6).

### *TLFM – comparable PL and RL performance, color tunability, and WL*

For the TLFM sample composed of *Per-C6-DCM* layers, the most complex system, we first investigated the influence of pumping power on RL performance and emission color. Since we are dealing with three dyes, we considered it valuable to excite the device in different ways. The middle layer is *C6*, while the outer layers are *Per* and *DCM* (see *Methods, Stacking configurations for TLFWM*). We decided to explore how the excitation side would affect the results. Initially, we excited the sample from the *DCM* layer side.

As seen in Fig. 3 (a), the required energy is significantly higher compared to SLFMs. This is due to the increased number of layers and dyes, which need to be excited to produce efficient laser emission.

**Figure 3: TLFM lasing performance, multicolor tunability, white lasing phenomenon, and device morphology.**

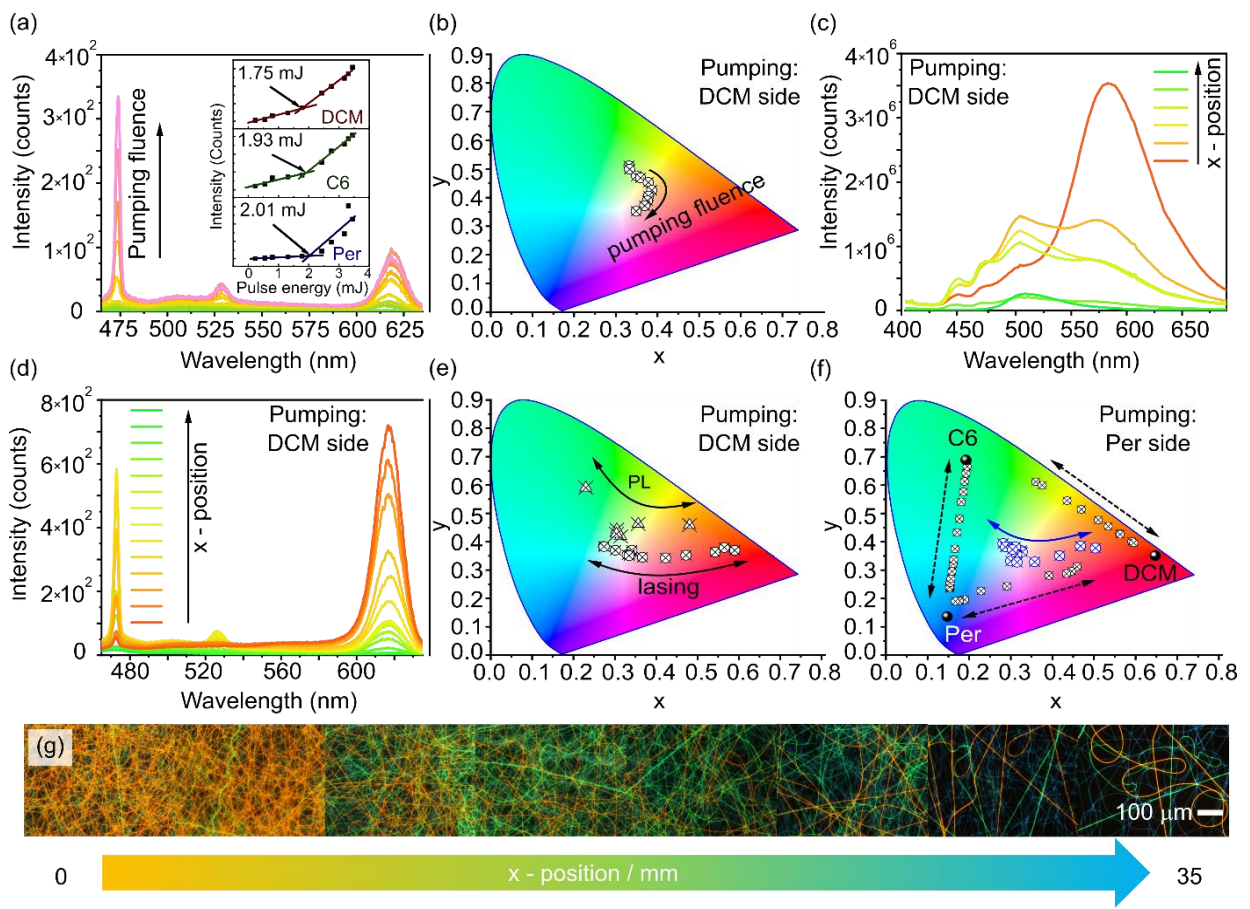

*Fig. 3* Per-C6-DCM TLFM emission spectra and lasing thresholds (a), CIE position evolution with pumping energy (b), PL color change with excitation position (c), RL spectra evolution with excitation position change (d), PL and RL color change with change of excitation position (e). All excitations from the DCM side (a-e). A comparison of color tuning ranges of studied samples showed that TLFM color tuning was obtained for excitation from Per side (f). A set of 8 microscopic images showing the morphological changes for different x-positions ranging from 0 to 35 mm with a 5 mm step (g).

The easiest to pump is the first layer containing *DCM* molecules. it requires an energy of 1.75 mJ/pulse (274kW). Next is the layer with *C6* dye with an energy threshold of 1.93 mJ/pulse (302kW). The hardest to pump is a layer with *Per* dye, showing the energy threshold of 2.01 mJ/pulse (3.15 kW). The details are demonstrated in section S6, Table S3. As can be observed, the threshold values are comparable, with no drastic differences. However, these slight divergences affect the color balance, visible in the CIE1931 space, as shown in Fig. 3 (b). At low pumping energy, the green emission from *C6* is most prominent. As the pumping energy increases, *DCM* begins to lase, and the collective emission shifts to yellow. Activation of the *C6* dye stabilizes the lasing color within the yellow range. Finally, when lasing from the *Per* dye is activated, the position on the CIE diagram changes direction, moving the emission point towards the desired white emission region, achieving a warm-toned white color. As seen, the emission color, in this case, can be controlled by adjusting the pumping energy, which influences the activation of different dyes and shifts the emission across the CIE1931 diagram.

Similar to the previous case of DLFMs, the TLFM sample also exhibits gradient architecture but with more complex fiber deposition densities (see Fig. 3 (c)-(g)). When only spontaneous emission is considered (Fig. 3 (c) and (e)), the excitation from one side of the device results in emission mostly from the *DCM* dye. As the x-position changes, the emission shifts toward a yellowish-green color, then incorporates a blue component (approaching white), and finally moves to pure green. In the case of RL excitation with 4 mJ/pulse, the lasing, initially dominated by the red *DCM* color, shifts towards warm white, following a path similar to temperature-based color tuning. Within an *x*-position range of approximately 8 *mm*, the RL remains nearly pure white. As the excitation moves further, the contribution of DCM to the emission becomes almost negligible, and the RL color transitions into lagoon green (see Fig. 3 (d) and (e)).

The morphology of the TLFM sample is difficult to analyze directly using a fluorescence microscope and Fiji software. The thresholding procedures fail to distinguish individual fibers, meaning a different method must be used. To gain insights into the fiber densities, we can instead analyze the changes in the intensity of different RL bands, which are well-separated. For a detailed analysis of the fiber deposition and the evolution of RL band intensities across different layers of the TLFM sample, please refer to section S7, Figure S7 in the ESI file.

As shown in Fig. S7, it is evident that the first Per layer is deposited roughly at the center of the microscopic glass substrate. The *C6* layer is then deposited on top of the *Per* layer, while the *DCM* fibers are shifted toward one edge of the substrate. Therefore, the RL color tuning ranges from DCM-dominated emission to a "blend" of *C6* and *Per*, with only a minor contribution from DCM.

The experiments presented so far were based on a configuration where we first excited the *DCM* fiber layer. Now, building on our previous results described in[46], we decided to excite the device from the *Per* side, expecting a more significant contribution from this blue dye. Indeed, as shown in Fig. 3 (f) on the CIE1931 diagram, the influence of *DCM* emission in the overall spectrum is less pronounced compared to the previous case. In the white color region, we observe a predominantly cold white emission, driven by the stronger contribution of the *Per* dye. This highlights that color tuning can be effectively controlled by adjusting the direction of device excitation, offering access to a wider range of colors, as discussed below.

**Discussion:**

In configuration, where the *DCM* layer is the first to absorb the incoming 405 nm pump light, it dominates the initial energy distribution. The pump light then passes through the central layer (with *C6*), where green lasing is observed, although it is slightly diminished due to the energy already absorbed by the red-emitting *DCM* layer. Finally, the *Per* dye absorbs residual light and provides blue lasing, which is essential for the collective, white lasing. Due to the higher contribution of the red-emitting *DCM* layer, the resulting white lasing exhibits a warmer hue. This shifts the color balance toward longer wavelengths, causing the light to appear close to the D55 standard (mid-morning / mid-afternoon daylight) and slightly less similar to D65 (noon daylight). The results depicting these findings are shown in the CIE XYZ 1931 diagram (section S8, Fig. S8), fragment (enlarged white light region). The red star in Fig. S8 represents the results obtained by exciting the device from the *DCM* side. The similarity (See Section S8, Table S4.) of the obtained white lasing coordinates to the D55 standard is 97.7% for the x coordinate and 97.5% for the y coordinate. Considering the calculated standard deviations, the similarity ranges are 95.0-99.6% for x and 94.0-99.1% for y. The narrow range of standard deviations indicates high reproducibility, underlining that the system produces stable white light with minimal color variation. This evidences strong control over the lasing color properties, which is crucial for applications requiring precise color calibration.

When pumping from the *Per* side (see the blue star in Fig. S8), the stacking sequence is reversed, and the cooler white hue observed is primarily due to the initial interaction of the pump light with the blue-emitting *Per* dye. This excitation order favors shorter wavelengths, resulting in a white lasing spectrum closer to the D70 and D75 standards (see Fig. S8), known for their bluish tones. The near-perfect resemblance to D70 indicates that the system can generate a highly accurate cool white hue with minimal discrepancies from this benchmark. The standard deviation ranges for D70 (96.0-99.8% for x and 96.9-99.3% for y) confirm high reproducibility in the cool-white emission. The similarity to D75 further supports the system's ability to produce light with bluish characteristics. In this case, the standard deviation ranges (93.8-98.5% for x and 94.6-99.7% for y) indicate a relatively tight grouping, although slightly more variation is observed compared to D70.

Finally, for both excitation configurations, the system still achieves excellent similarity to the widely used D65 standard. For pumping from the *DCM* side, the similarity is 96.1% and 97.1% for the x and y coordinates, respectively, with standard deviations ranging from 93.5-99.3%. When exciting *Per* first, the similarity improves to 97.9% and 98.4% (x, y), with a range of 96.0-99.2%. Compared to systems commonly referenced in the literature, it is shown that for organic, inorganic, and hybrid white lasers, discrepancies from the D65 standard may range around 8–10% for both coordinates [39,40]. For the y-coordinate, reported differences can be as low as 1.5%, while the x-coordinate shows a high variation exceeding 11% [38]. Moreover, the literature indicates that discrepancies can reach even a dozen percentage points, with some cases going beyond 20% in one coordinate [40,41]. On the contrary, our system remains within a tight tolerance for both D65 coordinates and achieves outstanding alignment with other widely used standards, including D55, D70, and D75. Table S4 presents the white lasing chromaticity coordinates (x,y) in the CIE XYZ 1931 color space obtained during the pumping of TLFM.

Multicolor laser systems should be evaluated also in terms of their RGB color range (See section S9, Figure S9). This assessment reflects their ability to cover specific portions of the color gamut. Good references are chromaticity models such as sRGB, AdobeRGB, or DCI-P3, which are critical for imaging technologies[47]. These three models are widely recognized color spaces, each tailored to specific applications in digital imaging, display technology, and spectroscopy[48]. sRGB, the most commonly used model, was designed for standard monitors and web content[49–51]. Covering about 35% of the visible spectrum, it prioritizes compatibility but has a relatively limited gamut. AdobeRGB expands the color range to approximately 50% of the visible spectrum, enabling more accurate reproduction of greens and cyans, making it ideal for professional photography, printing, and workflows requiring detailed color fidelity. DCI-P3, emerging from cinema standards and now integral to HDR and 4K displays, covers about 45.5% of visible colors, emphasizing vivid reds and greens to meet the demands of high-definition content. Below, the comparison of all mentioned models versus the RGB tunability achieved in this work is demonstrated (see Fig. S9).

Considering the green color, the fiber mats exhibit coordinates of (x = 0.2419, y = 0.6981), which closely approach the green defined in the demanding DCI-P3 color space (x = 0.265, y = 0.690). Also, SLFM's coordinates outperform those of sRGB (x = 0.300, y = 0.600) and AdobeRGB (x = 0.210, y = 0.710), indicating that the fiber mats would be well-suited for HDR and 4K light sources, which require a wide green gamut. For red, the SLFM's lasing is positioned at (x = 0.6296, y = 0.3695), which aligns closely with the red values of sRGB and AdobeRGB (both at x = 0.640, y = 0.330). This suggests that the red-emitting SLFM is well-suited for reproducing red tones in these two used color spaces. However, its red chromaticity does not extend to the more saturated red of DCI-P3 (x = 0.680, y = 0.320), which implies limitations for applications requiring this benchmark. The blue performance of the SLFM is more limited, with coordinates of (x = 0.1415, y = 0.1656) that differ significantly from the deeply saturated blue defined in sRGB, AdobeRGB, and DCI-P3 (all at x = 0.150, y = 0.060).

**Conclusions:**

To conclude, we have fabricated hierarchical fiber mats with gradients of fiber densities. This design was then utilized to demonstrate widely color-tunable PL and RL. The presented approach gives a wide possibility of laser emission color tuning, making it very versatile but still requiring further investigations. What is very promising is that there is still room for optimization, as, for example, different laser materials can be used instead of the ones described here. Electrospinning can be carried out for various luminescent materials, including π-conjugated polymers[52,53], colloidal quantum dots [54,55], TADF dyes [56], and many more. Therefore, their lasing properties can be improved further, especially to achieve lower threshold conditions.

The results underscore the high potential for precise color control. Through targeted design and fabrication, including the arrangement of electrospun layers with specific dyes, it is possible to create white lasers tailored to particular needs, offering control over the hue from warm to cool white within a compact, small-scale device. In addition to precise color tuning, this system shows great promise for applications in Li-Fi technology. By enabling efficient and tunable white-light emission, it can be integrated into Li-Fi systems, where optical communication relies on high-quality and coherent lighting. The ability to fine-tune the emission spectrum from cool to warm white also opens possibilities for adaptive lighting solutions, allowing for energy-efficient lighting that dynamically adjusts according to environmental or user preferences. This flexibility extends to laser display technologies currently emerging in the market.

**Methods:**

*Materials:*

The PMMA (Sigma Aldrich Mw ~350,000 by GPC) polymer solution in N,N-Dimethylformamide (DMF) (Sigma Aldrich, HPLC grade) of 15%*w/w* concentration was used for the electrospinning process for all investigated samples. To dope PMMA polymer with different laser dyes that can emit the light in red (R), green (G), and blue (B) regions of the visible spectrum, we have chosen the following dyes: Perylene from Sigma-Aldrich (B), Coumarin 6 (3-(2-benzothiazolyl)-7-(diethylamino)-2H-1-benzo-pyran-2-one) (G) and DCM ([2-[2-[4-(dimethylamino)phenyl]ethenyl]-6-methyl-4H-pyran-4-ylidene]-propanedinitrile) (R) both from Photonics Solutions Ltd. The concentration to PMMA mass was 3%.

*Electrospinning:*

All samples were prepared using a SpinBox (Fluidnatek) electrospinning system in the horizontal orientation of the nozzle and collector. For each sample, we used the flat collector and scanning emitter. The distance from the emitter in each case was 15 cm, the relative humidity (RH) was around 39-40%, and the temperature was 25 °C. The pump rate was always kept at 1.5 ml/h, and as the nozzle, we used a needle with a size of 27G. We have prepared 7 types of fiber mats. The voltage between the collector and emitter was always equal to 12 kV, with potentials equal to -4 kV and 8 kV applied to the collector and emitter, respectively. The scheme of the electrospinning process is shown in Fig. 1 (b).

In general, fabricated samples can be divided into three categories.

i) Single-layer fiber mats (SLFM), with only a single fiber layer containing one type of luminescent dye, spun for 3*min*.

ii) Double-layer fiber mats (DLFM) with a hierarchical, layer-by-layer structure, where one layer is spun for 1*min* and the second for 3*min*.

iii) Triple-layer fiber mats (TLFM), with hierarchical, layer-by-layer structure, where one layer is spun for 1*min*, the second one for 0.5*min,* and the last one for 3*min*.

The details about their fabrication processes can be found in Table S1 in ESI.

*Spectral characterization:*

Photoluminescence measurements were carried out using the FluoroMax4 spectrofluorometer from Horiba. Photoluminescence decay times were measured using a FLIM (Fluorescence Lifetime Imaging Microscopy) mode of Stellaris 8 confocal microscope from Leica.

*Random lasing measurements:*

RL measurements were carried out in an optical system composed of a Horizon I (Continuum) optical parametric oscillator (OPO) supplied by the third harmonic of a nanosecond Surelite II (Continuum) Nd:YAG laser, generating light of wavelength 405 nm passing through a broad-band waveplate, and a GaIn-Laser polarizer, used to set horizontal polarization and assure control over pulse energy. The shape of the pumping beam was adjusted using two diaphragms, the first inserted just behind the OPO exit port and the second placed before the polarizer. The sample was illuminated at an angle of 25 degrees concerning the sample's plane, and the emitted light was collected perpendicularly by SMA fiber with a 0.22 numerical aperture connected to an Andor Shamrock 500i spectrometer equipped with an Andor iDUS CCD camera. The spot size of the excited area had an elliptical shape of 5 and 5.5 mm axes; the excited area was 0.216 $cm^2$.

RL threshold conditions were estimated from RL intensity vs. pulse energy plots by finding the intersection point of two linear functions fitted to low-slope and high-slope parts of obtained intensity-to-pulse energy relationships.

Color tuning measurements of RL emission were carried out using the light of 4mJ of energy per pulse to ensure pumping over the threshold conditions.

*Stacking configurations for TLFWM*

Two experimental configurations were implemented to assess the impact of layer order on the resulting white lasing hue. In the first one, the sample was measured with the stacking sequence *DCM* (top), *C6* (central), and *Per* (bottom). In the second case, the measurement was taken from the opposite side, in the order of layers: *Per* (top), *C6* (central), and *DCM* (bottom). In both arrangements, the incident pump light initially encounters the top layer. For both experiments, nine and thirteen lasing points were collected within the white region, respectively. Based on the gathered data, the average values of the x and y coordinates were calculated to assess the similarity of the obtained values to the D-type coordinates, which are commonly accepted and used in the industry. The standard deviation of the sample measurements was also evaluated, as it reflects the variability of the results and provides insight into the consistency and reliability of the white lasing hue across the different experimental points.

*Morphological studies:*

Fiber-mat morphologies were obtained using a Stellaris 8 confocal microscope from Leica and a Scanning Electron Microscope (SEM) SM 6610LVnx from JEOL Ltd.

**Acknowledgments:**

L.S. would like to thank the National Science Centre of Poland for financial support with grant no. 2021/41/B/ST5/01797.

**Authors' contribution:** A.S., investigation (lasing, spectroscopy), methodology, formal analysis, data curation, original manuscript preparation, writing, reviewing, conceptualization; J.C., supervision; J.P., investigation (morphology – fluorescence microscopy), formal analysis, coding,; K.H.-S., samples preparation (electrospinning), investigation (morphology – fluorescence microscopy); K.K., investigation (morphology – SEM), reviewing; L.S., conceptualization, investigation (lasing), methodology, formal analysis, data curation, writing, reviewing, supervision, visualization, funding acquisition, and project administration; M.M., investigation (morphology – confocal microscopy).

**Data availability**

The datasets generated and/or analysed during the current study are available from the corresponding author on reasonable request.

# Supporting information

## *Polymer fiber mats provide white and full-color tunable lasing*

Alina Szukalska[1], Kinga Halicka-Stępień[1], Jastin Popławski[1], Katarzyna Kołodzińska[1], Michał Majkowski[2], Joanna Cabaj[1], Lech Sznitko[1]

[1] Wroclaw University of Science and Technology, Institute of Advanced Materials, Department of Chemistry, Wybrzeże Wyspiańskiego 27, 50-370 Wroclaw, Poland.

[2] University of Wroclaw, Faculty of Biotechnology, Plac Uniwersytecki 1, 50-137 Wroclaw, Poland

Corresponding authors: lech.sznitko@pwr.edu.pl, alina.szukalska@pwr.edu.pl

**Table of contents:**

## S1. Fiber's density studies

To determine the density of fiber distribution depending on spinning time, we decided to utilize the same PMMA solution with DCM dye described in the "Methods" section in the main body of the manuscript. Samples were prepared under the same conditions. The only variable was the spinning time, which was set to 30, 60, and 90 seconds. Photographs in Fig. S1 (a) show the global morphology of the obtained samples. The differences are clearly visible, indicating the highest number of fibers for the longest spinning time. After samples were fabricated, using a fluorescence microscope with the objective ×10, we collected a set of images (2592 ×1944 pixels), taken consecutively, one after another by scanning the sample, as presented in Fig. S1 (b). To each of the micro images, we adjusted the thresholding level to filter out the background from the fibers' fluorescence and binarized them. Each column (1×1944 pixels) of such transformed images was then analyzed according to the number of bright and dark pixels, where bright pixels indicated places occupied by fibers. When all photographs, after proper transformation, were stitched together, we were able to obtain histograms of "linear" fibers' densities understood in terms of the percentage of bright pixels collected in each image column. Thus, a single histogram box represented an area of size 1×1944 pixels with a step size of 1 pixel towards the scanning direction. Exemplary histograms with proper running averages are shown in Fig. S1 (c), indicating the complex spatial distribution of fibers over samples and the presence of density gradients.

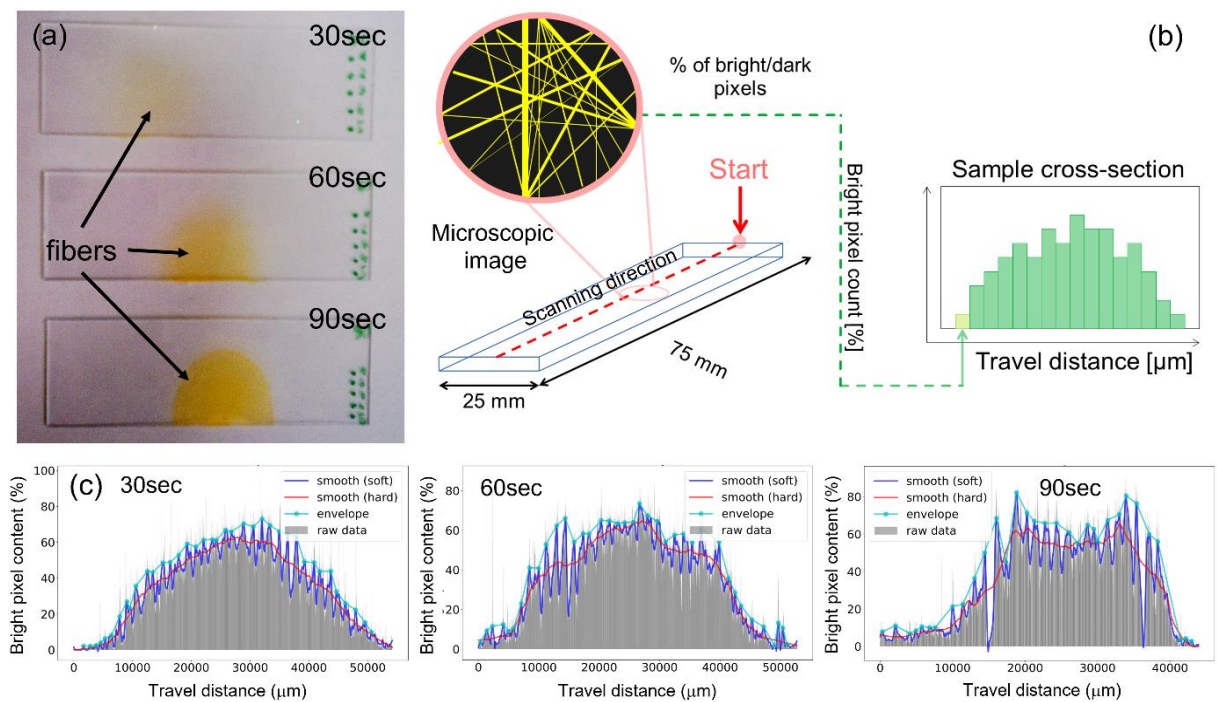

*Fig. S1.* *Sample's photographs (a), a scheme of protocol for generating fiber density histograms (b), and fiber density histograms for samples prepared using different spinning times (c).*

It can be seen that for 30 seconds of spinning, the higher amount of fibers is inside the sample, and the fiber coverage is gradually decreasing towards the sample's edges. For 60 seconds of spinning, the interior of the sample is nearly uniform, while the number of fibers decreases rapidly when a position is close to the sample's edge. For 90 seconds of spinning, the situation is somewhat different; the sample's interior is uniform, but the number of fibers close to the edges is, in this case, higher than in the central part. Thus, it is possible to obtain different fiber gradients by only changing the time of the electrospinning process.

a) Python code for images import, sorting, and thresholding:

```
#1.import all the images and sort them
#2.threshold each of them
#3.count the amount of white pixels in each column of each frame
#4.repeat over all frames
#5.export histogram of obtained statistics

#---------------------------------------------------------------------------

#importing libraries
import matplotlib.pyplot as plt
import cv2
import glob
import os
import time
import csv
import numpy as np
import skimage as sk
import tkinter as tk
from tkinter.filedialog import askdirectory

#importing files
root = tk.Tk()
root.withdraw()

file_path = askdirectory()
files = glob.glob(file_path + '/*.tif', recursive=True) or glob.glob(file_path + '/*.png', recursive=True)
#import images as either .png or .tif
files.sort(key=os.path.getmtime, reverse=False) #enforcing sorting in chronological order, which corresponds to location on the sample

#Script runtime measurement
start_time = time.time() #runtime estimation beginning

#processing the data
ratio = 1.77 #pixel to metric conversion ratio - depends on microscope / objective.
count = 1 #initial counter
y = []
```

```python
for file in files:
    image = plt.imread(file) #load initial image
    gray = cv2.cvtColor(image, cv2.COLOR_BGR2GRAY) #convert from RGB to grayscale
    histogram = gray.ravel() #compose histogram of image
    threshold = sk.filters.threshold_otsu(histogram) #histogram based thresholding using Otsu's method (note that colors are inverse - bright areas will be black, and dark areas will be white)
    binarized = gray <= threshold #actual conversion from grayscale to binary

    for column in range(binarized.shape[1]): #loop over all the columns in particular image
        black = np.sum(binarized[:, column] == True) #count all the black pixels
        white = np.sum(binarized[:, column] == False) #count all the pixels that are white (note the inversion of colors)
        percentage = (white / (white + black)) * 100
        count += 1 #increment the count of columns - number of iterations is equal to number of columns in all the processed images
        y.append(percentage)

x = range(1, count) #generate the x vector based on accumulated operation count
x = [i / ratio for i in x] #list comprehension to recalculate pixels into µm

#exporting the numerical data to .csv file for further analysis
with open(f'{file_path.split("/")[-2]}.csv', 'w') as file:
    writer = csv.writer(file)
    writer.writerow(['x', 'y'])
    for i in range(len(x)):
        writer.writerow([x[i], y[i]])

#return runtime - please note that this part only serves the feedback
end_time = time.time()
execution_time = end_time - start_time
print(f"Finished in: {execution_time:.2f} seconds")
```

    b)    Python code for data visualization:

```python
#plotting the output
import numpy as np
from matplotlib import pyplot as plt
from scipy.optimize import curve_fit
import scipy.signal as sig
import tkinter as tk
from tkinter.filedialog import askdirectory

#%% file path
root = tk.Tk()
root.withdraw()

file_path = tk.filedialog.askopenfilename()
```

```
x, y = np.loadtxt(file_path, skiprows=1, delimiter=',', unpack=True)

#smooth
smooth = sig.savgol_filter(y, 1250, 2) #soft smoothing of data
smooth_2 = sig.savgol_filter(y, 12500, 2) #hard smoothing of data
peaks, _ = sig.find_peaks(smooth, distance=500, prominence=0.5, width=500) #peak finding for data envelope fit

#plotting
plt.figure(figsize=(16,9))
plt.rcParams['font.size'] = 20
plt.bar(x, y, lw=1, color='k', alpha=0.3) #raw data
plt.plot(x, smooth, '-', c='b', lw=3.5, alpha=0.7)
plt.plot(x, smooth_2, '-', c='r', lw=3.5, alpha=0.7)
plt.plot(x[peaks], smooth[peaks], '-o', c='c', lw=3.5, markersize=11, alpha=0.7)
plt.title(f'{file_path.split("/")[-1][:-4]}')
plt.xlabel("Travel distance (µm)")
plt.ylabel("Bright pixel content (%)")
plt.legend(['smooth (soft)', 'smooth (hard)', 'envelope', 'raw data'], loc='upper right')
plt.savefig(f'{file_path.split("/")[-1][:-4]}.png', dpi=600, format='png') #plot export - if other format is required, just substitute "png" in name and format
#plt.show()
print("Done!")
```

## S2. Parameters for electrospinning

*Table S1.* *Electrospinning parameters for the obtained samples. Note that layer "A" is spun as the first layer, layer "B" as the second, and layer "C" as the last.*

| Sample name | Sample type | Layer | Dye | Spinning time |
|---|---|---|---|---|
| **DCM** | SLFM | A | DCM | 3 min |
| **C6** | SLFM | A | Coumarin 6 (Coumarin 540) | 3 min |
| **Per** | SLFM | A | Perylene | 3 min |
| **Per-DCM** | DLFM | A | Perylene | 3 min |
| | | B | DCM | 1 min |
| **DCM-C6** | DLFM | A | DCM | 1 min |
| | | B | Coumarin 6 (Coumarin 540) | 0.5 min |
| **Per-C6** | DLFM | A | Perylene | 3 min |
| | | B | Coumarin 6 (Coumarin 540) | 0.5 min |
| **Per-C6-DCM** | TLFM | A | Perylene | 3 min |
| | | B | Coumarin 6 (Coumarin 540) | 0.5 min |
| | | C | DCM | 1 min |

## S3. SLFM's lasing performance

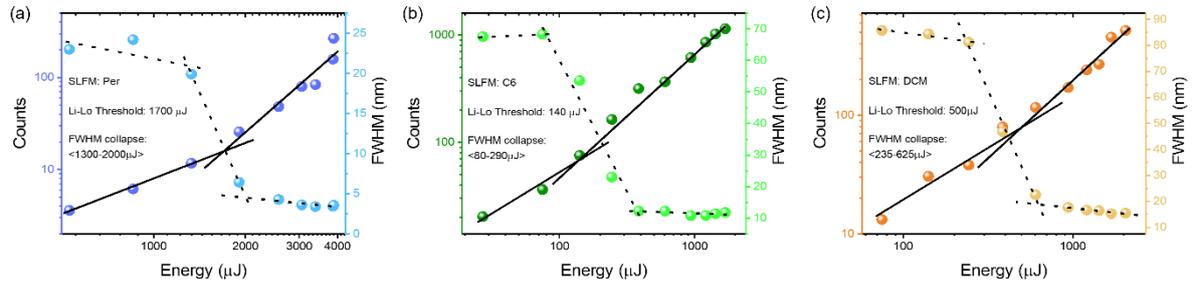

***Fig. S2.*** *Light in - light out (Li-Lo) and FWHM emission plots for SLFMs doped with (a) Per, (b) C6, and (c) DCM.*

*Table S2. Energy, peak power, energy fluence, and peak intensity threshold estimation for SLFMs.*

| Dye: | Per | C6 | DCM |
|---|---|---|---|
| **Li-Lo threshold** ($\mu$J/kW/mJcm$^{-2*}$/kWcm$^{-2*}$) | 1700/266/7.9/1230 | 140/22/0.7/102 | 500/78/2.3/361 |
| **FWHM collapse central position** ($\mu$J/kW/mJcm$^{-2*}$/kWcm$^{-2*}$) | 1650/259/7.6/1199 | 225/35/1.0/162 | 430/67/2.0/310 |
| **FWHM collapse range** | 1300-2000 $\mu$J<br>204-313 kW<br>6.0-9.3 mJcm$^{-2*}$<br>944-1449 kWcm$^{-2*}$ | 80-370 $\mu$J<br>13-58 kW<br>0.4-1.7 mJcm$^{-2*}$<br>60-269 kWcm$^{-2*}$ | 235-625 $\mu$J<br>37-98 kW<br>1.1-2.9 mJcm$^{-2*}$<br>171-454 kWcm$^{-2*}$ |

\* excited spot area $S = 0.216$ cm$^2$

## S4. FLIM and FRET estimation

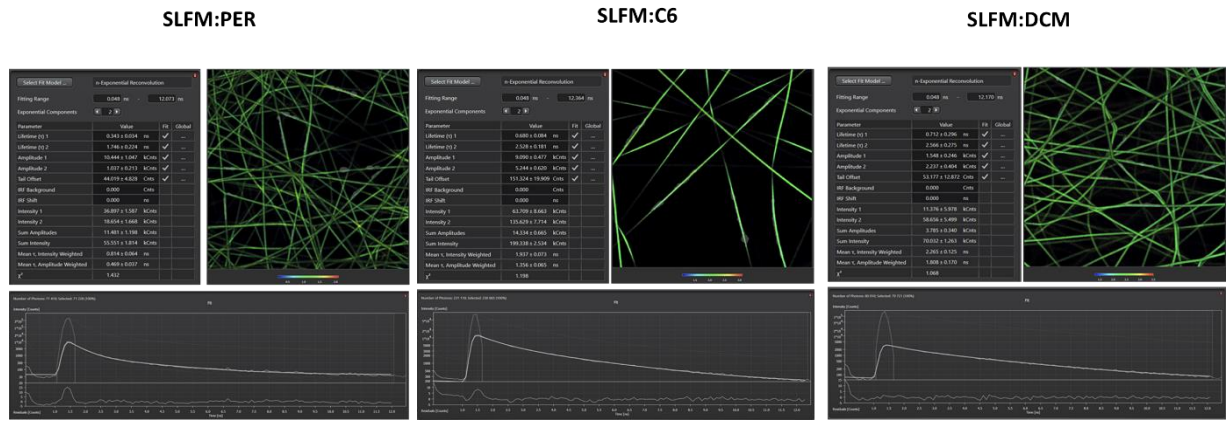

***Fig. S3.*** *FLIM fitting parameters for SLFM with different dopants.*

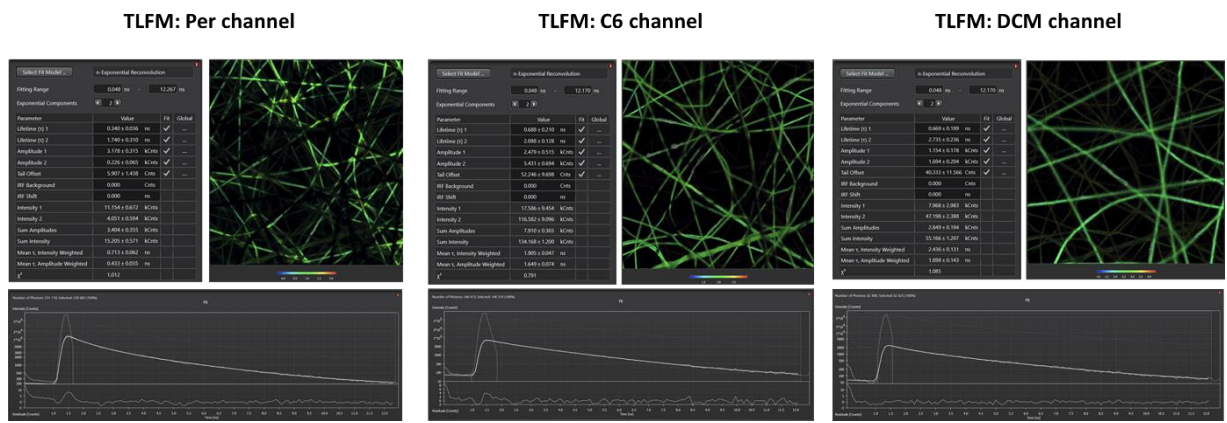

***Fig. S4.*** *FLIM fitting parameters for TLFM different dopant channels.*

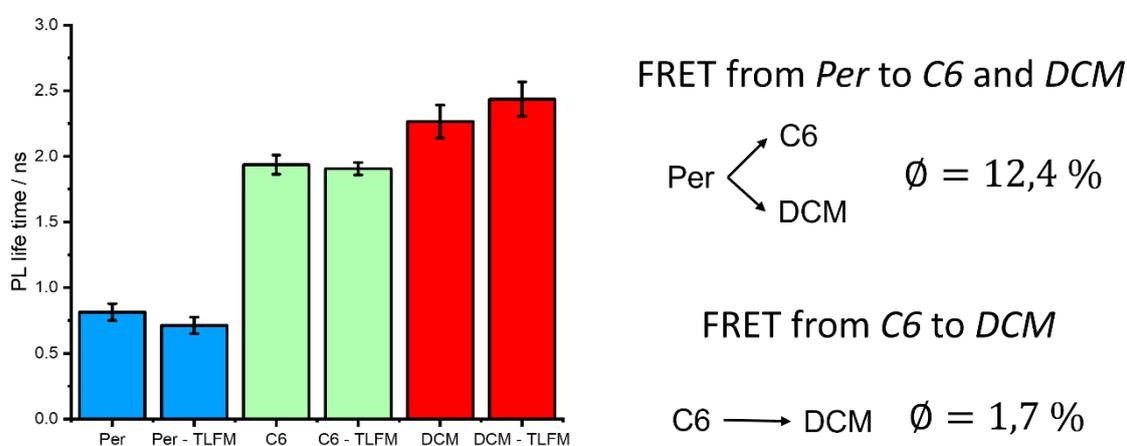

***Fig. S5.*** *PL lifetimes of different dyes in SLFM and TLFM with quenching efficiency of particular dye taken from the average lifetime.*

FRET estimated according to the equation: $\emptyset = \left(1 - \frac{\tau_{DA}}{\tau_D}\right) \times 100\%$, where $\tau_{DA}$ – an average lifetime of donor in presence of acceptor and $\tau_D$ – an average lifetime of donor without acceptor. As can be seen the influence of FRET effect is strongly reduced or even completely neglible if life-time estimation errors are considered.

## S5. DLFM fiber densities

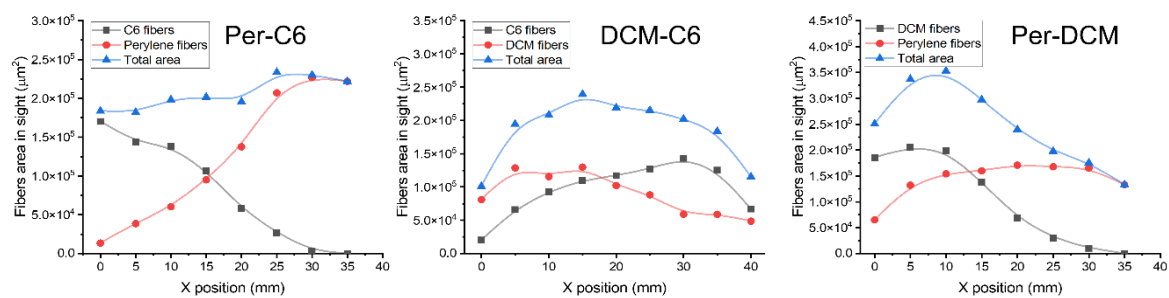

***Fig. S6.*** *Area covered by fibers in the area of observation under a fluorescence microscope for various DLFM samples: Per-C6 (left), DCM-C6 (central), and Per-DCM (right). The area of observation is $2.57\times10^5\,\mu m^2$.*

## S6. TLFM lasing performance

***Table S3****. Energy, peak power, energy fluence, and peak intensity threshold estimation for SLFMs.*

| Dye: | Per | C6 | DCM |
|---|---|---|---|
| Li-Lo threshold | | | |
| $\mu J$ | 2010 | 1930 | 1750 |
| kW | 315 | 302 | 274 |
| mJcm$^{-2*}$ | 9.3 | 8.9 | 8.1 |
| kWcm$^{-2*}$ | 1460 | 1400 | 1270 |

* excited spot area $S = 0.216\ cm^2$

## S7. RL intensity vs. x-position of different dyes in the TLFM sample

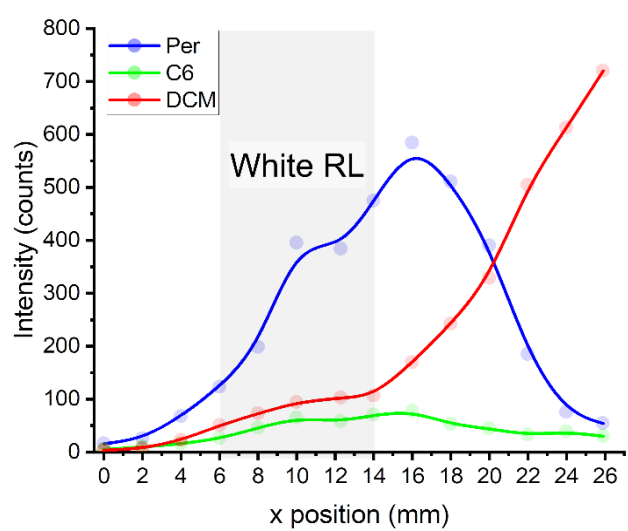

***Fig. S7.*** *The intensity of Per, C6, and DCM RL obtained from the TLFM sample with respect to different excitation positions.*

## S8. White color determination

Whiteness standards such as D50, D55, D65, D70, and D75 are widely accepted references in the color space, particularly in daylight illumination. They are defined by their correlated color temperatures and represent specific shades of white that are tailored for diverse scientific, artistic, and industrial applications. The D50 illuminant approximates natural light at a color temperature of 5000 K, offering a warm white that is adopted in printing and art-related color assessments. Its spectrum allows for accurate reproduction of colors, making it the preferred choice for environments that require precise visual evaluation. It is suitable for applications where neutral lighting is desired, such as in photographs taken during sunset.

D65 is the most widely recognized standard, matching midday daylight with a temperature of approximately 6500 K. It is commonly used in photography, graphic design, spectroscopy, imaging, and even medicine [1–3] since it best represents color perception under natural light. D70, with a color temperature of around 7000 K, is a cooler white used in applications requiring enhanced color accuracy in bright lighting conditions, such as laboratories. D75, with a color temperature of about 7500 K, represents cold daylight and is often utilized in specialized applications such as textile whiteness evaluation, where accurate color reproduction is crucial in challenging lighting conditions. Each standard allows precise color comparison and enhances product quality across various industries.

*Table S4. Comparison of the WL coordinates in the measured TLFM device versus the standard of the D series.*

| Standard Illuminant | x | y | % similarity (xExp to xD) | % similarity (yExp to yD) | The standard deviation range for % similarity (x) | The standard deviation range for % similarity (y) |
|---|---|---|---|---|---|---|
| **From the DCM side** | | | | | | |
| D50 | 0.34567 | 0.35850 | 94.0% | 94.5% | 91.4 – 96.6% | 91.2 – 97.7% |
| D55 | 0.33242 | 0.34743 | **97.7%** | **97.5%** | **95.0 – 99.6%** | **94.0 – 99.1%** |
| D65 | 0.31272 | 0.32903 | **96.1%** | **97.1%** | **93.3 – 99.0%** | **93.5 – 99.3%** |
| D70 | 0.3054 | 0.3216 | 93.6% | 94.7% | 90.7 – 96.6% | 91.0% - 98.4% |
| D75 | 0.29902 | 0.31485 | 91.4% | 92.4% | 88.4% – 94.4% | 88.7% - 96.2% |
| Experiment | 0.32483 | 0.33863 | | | | |
| **From the Perylene side** | | | | | | |
| D50 | 0.34567 | 0.35850 | 88.5% | 90.3% | 85.2 – 91.9% | 88.1 – 92.6% |
| D55 | 0.33242 | 0.34743 | 92.1% | 93.2% | 88.6 – 95.5% | 91.0 – 95.5% |
| D65 | 0.31272 | 0.32903 | **97.9%** | **98.4%** | **94.2 – 98.5%** | **96.0 – 99.2%** |
| D70 | 0.3054 | 0.3216 | **99.8%** | **99.3%** | **96.0 - 99.8%** | **96.9 - 99.3%** |
| D75 | 0.29902 | 0.31485 | **97.7%** | **97.1%** | **93.8 – 98.5%** | **94.6 – 99.7%** |
| Experiment | 0.30604 | 0.32389 | | | | |

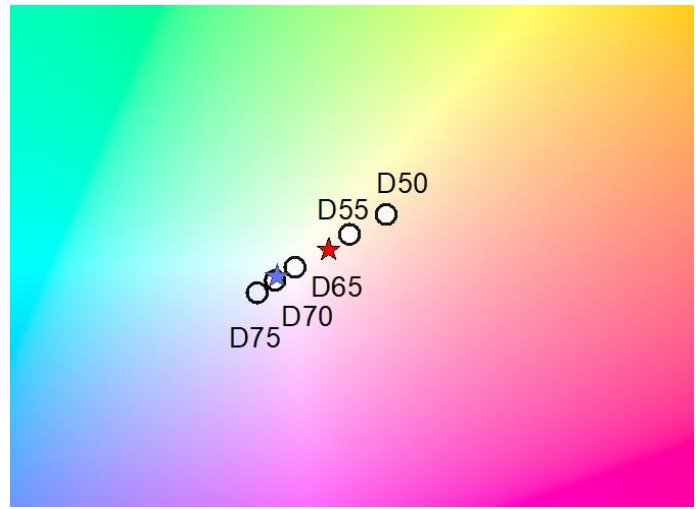

***Fig. S8.*** *Comparison of experimental WL coordinates vs. D-series standards in the CIE XYZ 1931 diagram. Red and blue stars mark the results for the measurements from the DCM and Perylene sides, respectively*

## S9. Color range tuning

***Table S5.*** *Coordinates (x, y) in the measured SLFMs compared to the sRGB, AdobeRGB, and DCI-P3 models.*

| Color | Fiber Mats | sRGB | AdobeRGB | DCI-P3 |
|---|---|---|---|---|
| Green | (0.242, 0.698) | (0.300, 0.600) | (0.210, 0.710) | (0.265, 0.690) |
| Red | (0.623, 0.370) | (0.640, 0.330) | (0.640, 0.330) | (0.680, 0.320) |
| Blue | (0.142, 0.066) | (0.150, 0.060) | (0.150, 0.060) | (0.150, 0.060) |

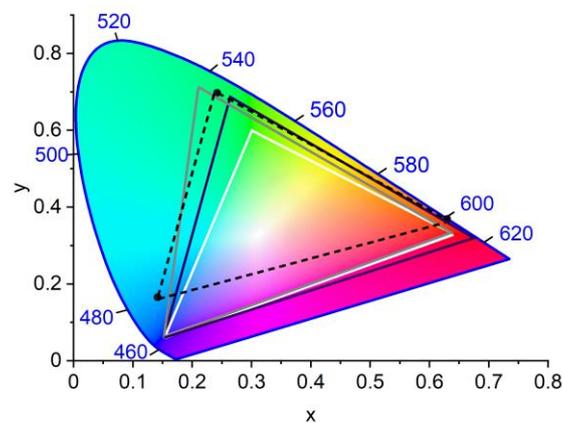

***Fig.S9.*** *Multicolor range obtained with red, green, and blue SLFMs (black, dashed lines) versus sRGB (white solid lines), Adobe-RGB (gray solid lines), and DCI-P3 (navy solid lines).*